\documentstyle[aps,twocolumn]{revtex}

\begin{document}
\draft
\title{ Temperature-dependent crossover in fractional quantum Hall edges in the presence of
 Coulomb interaction}
\author{Wenjun Zheng and Yue Yu}
\address{Institute of Theoretical Physics,  Academia Sinica,  Beijing 100080, 
China}

\date{}
\maketitle
\begin{abstract}

      Based on a newly derived microscopic fractional quantum Hall edge model, we study its
thermodynamics at finite temperature. For the dressed energy spectrum a critical
energy scale determined by the temperature exists, below which the refractive dispersion
which is essential to the model's bosonization to a Luttinger liquid is smeared. According to this observation,
a temperature-dependent crossover picture is proposed, and applied to the
analysis of a recent tunneling experiment in comparison with the standard Luttinger
liquid theory, and a better fit to the features of the measured conductance-temperature curve is achieved.
We also consider the role of the Coulomb interaction in the thermodynamics and find that a
crossover exists during which the influence of the Coulomb interaction is suppressed with the increase of temperature.

\end{abstract}

\pacs{PACS numbers:  73. 40. Hm, 71. 10. +x, 71. 27. +a}
Since the sensational discovery of fractional quantum Hall effect(FQHE) in 1982 \cite{Tsui},
many efforts have been devoted to the understanding of this fascinating 2D strongly
correlated system \cite{Stone}. In the theoretical aspect, progress has
been made to account for its low energy physics which is determined by the FQHE edge
states because of the gapful bulk excitations. Compared with the integer quantum Hall edge states which are well
understood in a chiral Fermi liquid picture\cite{Halp}, the description of the FQHE edge
 is much more subtle. There is still some controversy in the real effects of electron correlations.
 One of the most concerned
questions is whether the fractional edge is a Fermi liquid (FL)or a Luttinger
liquid(LL)\cite{LF}. From the viewpoint of the macroscopic effective theory, Wen
advanced the chiral Luttinger liquid (CLL) picture \cite{Wen}, which emphasizes the
non-Fermi liquid nature of the FQHE edge states due to strong correlations. Some
predictions \cite{LL1}\cite{LL2} based on CLL have been corroborated in a recent
tunneling experiment between $\nu=1/3$ fractional edges\cite{TUN}. Meanwhile, a
microscopic composite fermion (CF) \cite{Jain} picture was used to construct a mean
field level edge model, which is basically a Fermi liquid one\cite{KW}. It is not surprising
that this naive mean field theory also found experimental supports from the
measurement of the anti-dot AB effect \cite{AB} and the resonant tunneling width \cite{GM}, for its bulk counterpart has already achieved impressive experimental successes.
In a recent study \cite{YZ1}\cite{ZWJ}, we succeed in relating the above two pictures
with each other by giving a microscopic derivation of the CLL model\cite{PM}. In our
derivation, we perform a composite fermion transformation on the electron
Hamiltonian, which attaches an even number of  flux quanta onto the electrons to form 
 the composite particles. While the statistical gauge field interactions between the edge
 particles and the bulk ones are treated at a mean field level, the gauge interactions between
the edge particles are taken care of explicitly, which lead to the square inverse interaction
 characteristic of the Calogero Sutherland model(CSM)\cite{CS}. In this way, we can 
 show that the FQHE edge states are described by the CSM with the residue interactions between the CFs treated as
perturbations. Then by means of bosonization and consideration of the chirality
, we finally arrive at Wen's CLL theory. The role of the residue short-range interactions is
also considered and proven to be unable to change the critical exponents of the CSM 
because of the chirality\cite{IGN}.
 In this letter, we are going to reconcile
the conflicts between the Fermi liquid and the Luttinger liquid descriptions by studying
the thermodynamics of the present microscopic edge model at finite temperature. We argue
that the refractive dispersion of the dressed energy which is essential to the CSM's bosonization to
a LL is smeared in the "window" opening around the Fermi point, whose width is
set by the temperature. The physical processes involving only excitations of
much lower energy scales will show FL like behaviors, while those including higher energy
 excitations will undertake a crossover to LL with the increase of energy scales. This mechanism\cite{MEC} makes it possible to
reconcile between the experimentally detected LL behaviors and the FL ones. We then apply
this intuitive crossover picture to the analysis of a recent tunneling experiment from
a normal metal to the $\nu=1/3$ fractional edge, and succeed in explaining two features of
the $G-T$ data which are inconsistent with the standard LL prediction. We also
discuss the role of the Coulomb interaction in the thermodynamics, and argue that finite
temperatures tend to suppress its effects.

     Before starting with the discussion of the thermodynamics, we first give a concise review of
the microscopic edge model. Both the thermodynamics and the bosonization
of it stem from the well-known thermodynamic Bethe ansatz (TBA)equation\cite{YY} that
reads
\begin{eqnarray} \epsilon(k)=\epsilon_0(k)+\frac{T}{2\pi}\int\limits_{-\infty}^{\infty}
\Phi^{\prime}(k-k^{\prime}) \ln(1+e^{-\frac{\epsilon(k^{\prime})}{T}}) dk^{\prime}
\end{eqnarray} 
where $\epsilon(k)$ is the dressed energy, $\epsilon_0(k)$ stands for
the bare energy given by $ \epsilon_0(k)=k^2-\mu $, $\mu$ is the chemical potential
, $\Phi(k)$ represents the phase
shift function of the model interaction, which in the CSM reads  $
\Phi_{CS}(k)=(\lambda-1)\pi sgn(k) $, and the model interaction is $
V_{CS}(x)=\frac{\lambda(\lambda-1)}{4} \biggl[\sin(\frac{x}{2})\biggr]^{-2} $. The
free energy at temperature $T$ is given by the dressed energy as follows,
\begin{eqnarray}
F&=&U-TS \\  \nonumber
&=&-T\int\limits_{-\infty}^{\infty} \rho_0 \ln(1+e^{-\frac{\epsilon(k)}{T}}) dk 
\end{eqnarray}
where $\rho_0=\frac{1}{2\pi}$.
 In order to bosonize the low energy effective
theory of the CSM, one goes to the zero temperature limit\cite{Wu} , and the TBA
equation reduces to $$
\epsilon(k)=\epsilon_0(k)-\frac{1}{2\pi}\int\limits_{-k_{F}}^{k_{F}}\Phi^{\prime}(k-k^{\prime})
\epsilon(k^{\prime}) $$ and one obtains the following refractive dispersion of the
dressed energy 
\begin{equation} \epsilon_\pm (k)=\Biggl\{ {\begin{array}{ll} \pm
v_F(k\mp k_F)+k_F^2, \;\;\; &|k|>k_F, \\ \pm v_F (k\mp k_F)/\lambda+k_F^2, &|k|<k_F.
\\ 
\end{array}}
\end{equation}
where $v_F$,$k_F$ are the Fermi velocity and the Fermi momentum respectively.
  Based on this dispersion relation and the ground state
energy formula $ E=\int\limits_{-k_F}^{k_F} \rho_0 \epsilon(k) dk $, one can bosonize
the CSM and arrive at the low energy fixed point of LL, with the characteristic
exponent $g=\lambda$. We emphasize the importance of the refractive dispersion,
which is essential to the conclusion of non-Fermi liquid($g\not=1$) in the CLL picture.

Then we turn to the situation of finite temperature, and calculate the dispersion of
the dressed energy near the Fermi point. A momentum scale $k_T$ set by $T$ is given by
$k_T=T/v_F$. For $|k-k_F|<<k_T$, or $|\epsilon(k)|<<T$, the TBA equation reduces
to 
\begin{equation} 
\epsilon(k)=\epsilon_0(k)+T(\lambda-1) \ln
(1+e^{-\frac{\epsilon(k)}{T}}) 
\end{equation}
One can expand it to the
$O(\frac{\epsilon(k)}{T})$ order, and obtain
\begin{equation}
\label{NON}
\epsilon(k)=\frac{2}{1+\lambda}\epsilon_0(k)+const+ O[(\frac{\epsilon(k)}{T})^2)]
\end{equation}
where the temperature-dependent part of $\epsilon(k)$ is absorbed into
the chemical potential to satisfy the condition $\epsilon(k_F)=0$, with the remaining
temperature-independent term $\frac{2}{1+\lambda}\epsilon_0(k)$ still meaningful as a physical energy. One can see that the
refractive dispersion at $T=0$ is smeared near the Fermi point at finite
temperature. Then we consider the situation of $|k-k_F|>>k_T$, or
$|\epsilon(k)|>>T$. One can easily obtain the dispersion relation similar to that of
zero temperature as follows
\begin{equation} 
\epsilon_\pm (k)=\Biggl\{ {\begin{array}{ll}
\epsilon_0(k)+O(e^{\frac{-|\epsilon(k)|}{T}}), \;\;\; &|k|>k_F, \\
\epsilon_0(k)/\lambda+O(e^{\frac{-|\epsilon(k)|}{T}}), &|k|<k_F.  \\ \end{array}}
\end{equation} 
where the refractive dispersion is restored. Therefore the dispersion
relation of the dressed energy that is essential to the CSM's bosonization to a LL is smeared at energy
scales lower than $T$, where the apparent "refractive index" turns from $\lambda$ to 1,
and a Fermi liquid like low energy effective theory is obtained.  
In this way both FL and LL behaviors can be accommodated at finite $T$, depending on the energy scales involved in
different physical processes.

Let us first discuss the equilibrium thermodynamics,
which is determined by the free energy $F$ in the canonical ensemble formulation , and $F$ 
can be divided into the sum of two parts as follows 
\begin{eqnarray}
F=F_l+F_h
\\  \nonumber
F_l&=&-T\sum_{r=-1,1}\int\limits_{|k-rk_F|<k_T}^{} \rho_0
\ln(1+e^{-\frac{\epsilon(k)}{T}}) \\ \nonumber
F_h&=&-T\sum_{r=-1,1}\int\limits_{|k-rk_F|>k_T}^{} \rho_0
\ln(1+e^{-\frac{\epsilon(k)}{T}})    \nonumber
\end{eqnarray} 
and $F_l$ is more important because
low energy excitations contribute more to the temperature-dependent part of $F$ which
determines the thermodynamics. One can see that $F_l$ contains the contribution from
the FL like dressed energy spectrum at low energy scales, while $F_h$ contains the
contribution from the LL related refractive spectrum at large energy scales. So the
thermodynamic functions derived from $F$ have mixed properties of FL and LL. Attempts
to clarify the FL/LL ambiguity, or to extract the corresponding characteristic
parameters (such as $v_F$,$v_N$ of LL) by measuring the equilibrium thermodynamic
quantities are expected to be improper. However, experiments designed to probe 
predominantly low energy excitations or high energy ones (relative to $T$) should be
able to discriminate between FL and LL behaviors, such as the measurement of
the temperature and the bias voltage dependences of the tunneling conductance $G(V,T)$. We
first consider the situation of tunneling from a normal metal to a 1/3 fractional
edge at fixed $T$. Since $eV$ corresponds to the energy scale of the most relevant
excitations probed by the measurement of $G(T,V)$, one can expect the behavior of
$G(T,V)$ to undertake a crossover from FL to LL with the increase of $V$. For $eV
<< T$ ,the $V$ independent behavior is expected, which results from the FL like low
energy excitations; while for $eV>>T$ the power law property of LL is restored,
that is 
\begin{equation}
G(T,V)\propto\Biggl\{ {\begin{array}{ll} V^{2}, \;\;\;
&eV>>T, \\ V^0, &eV<<T.  \\ 
\end{array}} 
\end{equation} 
To verify the above
qualitative prediction, we cite the experimental results of Chang et al.\cite{CPW} for comparison.
In Fig.1a, the measured $I-V$ relation at fixed $T$ experiences a crossover from
$I\propto V$ to $I\propto V^3$ near $eV\approx T$, which means a corresponding
crossover for $G(T,V)$ from $V^0$ to $V^2$, as predicted above. The quantitative
result of the $I-V$ curve based on the crossover picture is shown in Fig.1b for
comparison. One can see that the fit is satisfactory. However, a good fit to the 
experiments can also be achieved by means of the scaling relation derived from
the standard LL theory. In order to demonstrate that the current crossover picture
is beyond an equivalent of the LL theory, we will turn to the situation of
variable $T$, where the present picture is able to describe more features of the
measured $G-T$ data than the standard LL theory.

   To make a quantitative comparison between the prediction of the crossover picture
and that of the LL theory, we first cite the basic results of the latter. By taking both $T$
and $V$ dependences into account, Kane and Fisher suggest a universal scaling form for
the $I-V$ relation that holds in the limit $G_{tun}<<G_{Hall}=e^2/3h$ (suppose
$\nu=1/3$). For tunneling into the $1/3$ edge from a normal metal, it is given by
\cite{KF} 
\begin{equation} 
I\propto T^\alpha[x+x^\alpha],  
\end{equation}
where $x=\frac{eV}{2\pi T}$, $\alpha=3$.   
Therefore $G(T,V)$ can be written as
\begin{equation}
G(T,V) \propto T^2+BV^2
\label{EQ} 
\end{equation}
where $B$ is a constant.
We then cite the
experimental results of reference\cite{CPW} in Fig.2a, which will be the bases of
the following discussions and comparisons. After a careful examination of Fig.2a, one
can discover two features that are inconsistent with eq(\ref{EQ}), but can be understood
qualitatively within the crossover picture. First, The two curves of $G(T,V)$ tend
to saturate with the increase of $T$, before the one with lower
$V$ warps up (the so called meander
behavior in ref\cite{CPW}), while the other with higher $V$ continues to saturate
 when approaching the gapful hole excitations. That means the exponent of $T$ tends to decrease with the
 increase of $T$. However, according to eq(\ref{EQ}), the exponent should
go up to 2 with the $V^2$ term being suppressed by larger $T^2$. This feature
can be understood in the crossover picture. When we
increase $T$ while fix $V$, more and more relevant excitations belong to the FL like part of the spectrum, which has
the effect of gradually lowering the exponent from the LL value 2 to the FL value 0, although the
crossover is disrupted by the gapful excitations. Secondly, we note that the
$G(T,V)$ curve with larger $V$ has larger exponent\cite{EXP3}. But eq(\ref{EQ})
predicts the opposite
 result, since larger $V^2$ weakens the power law behavior of
$T^2$. This again can be explained qualitatively here, because larger $V$ means to probe
higher energy excitations which tend to reach the LL region of the dressed energy
spectrum, and as a result leads to larger exponent.

    In order to give a quantitative justification of the above argument,
we calculate $G(T,V)$ by
using the following tunneling density of states
\begin{equation} N(E)\propto\Biggl\{
{\begin{array}{ll} N_0, \;\;\;&E<<T, \\ E^2, &E>>T.  \\ 
\end{array}} 
\end{equation}
$N(E)$, for $E$ near the order of $T$, corresponds to a crossover between the above two
limiting regions\cite{CO}. $N_0$ does not depend on $T$ because $\frac{2}{1+\lambda}\epsilon_0(k)$ in eq(\ref{NON})is
 $T$-independent .
 $G(T,V)$ is given by 
$$ 
G(T,V)\propto
\int\limits_{-\infty}^{\infty}[f^{\prime}(E+eV)+f^{\prime}(E-eV)](1-f(E))N(E)dE
$$
where $f(E)=1/(1+e^{E/T})$ is the Fermi distribution function.
The numerical results are shown in Fig2.b. The basic features of Fig.2a are reproduced
satisfactorily. Even the warping up behavior is manifest, which can be analyzed as
follows. According to the expression of $N(E)$, the contributions of the LL part and the FL like part can be separated in
$G(T,V)$ as $$ G(T,V)=W_{FL}(V/T)+W_{LL}(V/T)T^2 $$ where $W_{FL}(x)$ and $W_{LL}(x)$
are weight functions of the two contributions. With the increase of $T$ from the order
of $V$, the variations of weight functions at first dominate the $G-T$ behavior by
transferring weight from $W_{LL}$ to $W_{FL}$, which leads to the slight saturation
of $G(T,V)$. However, if we further increase $T$, the weight functions themselves tend to
saturate, while the $T^2$ term begins to dominate the $G-T$ behavior, which results in
the warping up tendency. With the further increase of $T$, the exponent rapidly approaches
the LL value 2. For large enouch $V$, the warping up behavior may be suppressed by the
approaching of the gapful excitations which tend to saturate $G_{tun}$ to
$G_{Hall}=e^2/3h$. This explains the loss of meander in the experimental curve with
larger $V$. Therefore, the FL behavior of $G(T,V)$ only manifests itself in a small interval of $T$
on the condition that $V$ is small enough, although the contribution
of the FL like excitations persists at the high temperature limit. The detailed fit between
theory and experiments will be reported elsewhere\cite{ZY}.

  From the above discussions and comparisons, one can see that a
smooth crossover from LL to FL like behaviors occurs with the increase of $T$. If we take into
account the finite size effects, the crossover temperature $T_c$ is roughly
determined by the length scale $L$ as $ T_c=v_F \frac{h}{L} $. For $T<T_c$, the
FL window near the Fermi point is actually absent, while for $T>T_c$, the FL like
effects begin to show up. In this way, both the finite temperature effects and the
finite size effects are coherently related, and they
are complementary to each other in that whenever one becomes weaker the other
will turn stronger.In typical experiments on the FQHE
edge tunnelings in a mesoscopic sample, $T_c$ can be as large as the order of $T$,
and one can only measure a mixed property of FL and LL. For instance, in
ref\cite{CPW} where the tunneling into the $\nu=1/3$ fractional edge is probed, the
measured exponents ($\alpha-1=1.75,1.5$) in the $G-T$ relation stay between the
 LL value 2 and the FL value 0. Therefore, a quantitative 
understanding of the crossover region in such as $N(E)$ for $E$ near $T$ is necessary, before one can
explain the transport experiments on the fractional edges explicitly. This will be
 studied in our future work.

   We then turn to the discussion of the role of the Coulomb interaction at
finite temperature. In a recent paper\cite{ZWJ}, we argue that
the perturbative interactions between the CFs with their force-range shorter than $1/x^2$ can
not renormalize the exponent $g=1/\lambda$, and the bosonization process based on the zero
temperature limit of TBA is still useful. However, these arguments are not
applicable to the long-range Coulomb interaction, since its phase shift function near
$k=0$ is even more singular than $\Phi_{CS}(k)$, and can change the long wave
length excitations of the CSM substantially. To see what happens when $T$ is not zero, 
 let us go back to the TBA equation, where the only change is in $\Phi(k)$, which is
modified by the introduction of the Coulomb interaction\cite{EXP2}.
\begin{eqnarray}
\Phi(k)&=&\Phi_{CS}(k) +\Phi_{sig}(k) \\
       &=&\Phi_{CS}(k) +C_1\frac{1}{k/k_C}+C_2\frac{\ln(k/k_C)}{k/k_C}
\end{eqnarray}   
   where $C_1$ and $C_2$ are two constants, and $k_C$ is the momentum scale set
by the Coulomb interaction:$k_C=1/a_B$, with $a_B$ the corresponding Bohr radius.

   With the inclusion of $\Phi_{sig}(k)$, one can rederive the same dispersion
relation at finite $T$, if the condition $ k_T>>k_C $ is satisfied. The key to this
invariance is the following two formulae : 
$\int\limits_{-\infty}^{\infty}\Phi_{sig}^{\prime}(k) dk=0$, and
$\int\limits_{-\infty}^{\infty}\Phi_{sig}^{\prime}(k)k dk=0$. Therefore, one can expect a
crossover from the Coulomb relevant regime to the Coulomb irrelevant regime with the
increase of $T$. This helps to explain why the influence of the Coulomb interaction is not
obvious in experimental observations at finite $T$, while it may change the
ground state properties of CLL considerably.

   In conclusion, we propose a temperature-dependent crossover picture based on the calculation of
the dressed energy at finite temperature, and then apply it to the
analysis of a recent tunneling experiment. We arrive at a fit to the data better
than that of the LL theory. We also discuss the effects of finite temperature on
the Coulomb interaction, and find a crossover behavior with the increase of temperature.

  One of us(WJZ) is grateful to Hua-Gang Yan for his help in the numerical calculations.
We also acknowledge the stimulating discussion with Q.Niu. This work is partly
supported by the NSF of China.


\begin{figure}
\caption{ (a)$I-V$ characteristics for tunneling from the bulk-doped n+ GaAs into
the $\nu=1/3$ fractional edge in a log-log plot for sample 1 at $B=13.4T$(crosses),
 and samples 2 at $B=10.8T$(solid circles). The solid curves represent fits to the
LL scaling form for $\alpha$=2.75 and 2.65, respectively[22]. 
(b) numerical result of $I-V$ log-log plot. The diamonds represent data generated by numerical integral based on
 the crossover picture. }
\label{f1}
\end{figure}

\begin{figure}
\caption{ (a)Log-log plot of the temperature dependence of $G(T,V)$ at low voltage
 bias for samples 1(upper curve) and samples 2(lower curve) at $\nu=1/3$. 
 The respective voltage biases are 4.97 and 2.64 $\mu V$. The solid straight lines 
 represent power laws with the respective exponents $\alpha-1$ of 1.75  and 1.5 [22]. 
 (b)numerical result of $G(T,V)-T$ log-log plot based on the crossover picture, the two curves correspond to 3.3$\mu V$(circles)
  and 2.1$\mu V$(diamonds), respectively. }
\label{f2}
\end{figure}
          

\begin{references}

\bibitem{Tsui}  D.C.Tsui, H.L.Stormer and A.C.Crossard, Phys. Rev. Lett.{\bf 48}, 1559 (1982). 
\bibitem{Stone} M. Stone, Quantum Hall Effect( World Scientific, Singapore, 1992).
\bibitem{Halp} B. I. Halperin, Phys. Rev. B{\bf 25}, 2185(1982).
\bibitem{LF}   M. R. Geller, D. Loss, G. Kirczenow, Phys. Rev .Lett.{\bf 77}, 5110(1996).
\bibitem{Wen}  X. G. Wen, Phys. Rev.Lett.{\bf 64}, 2006(1990); Phys. Rev. B{\bf 41},
 12838 (1990).
\bibitem{LL1}  K. Moon, H. Yi, C. L. Kane, S. M. Girvin, and M. P. A. Fisher,
Phys. Rev. Lett.{\bf 71}, 4381(1993).
\bibitem{LL2}  P. Fendley, A. W. W. Ludwig, and H. Saleur, Phys. Rev. lett.{\bf 74}, 3005(1995).
\bibitem{TUN}  F. P. Milliken, C. P. Umbach, and R. A. Webb, Solid State Commun
{\bf 97}, 309(1996).  
\bibitem{Jain} J. K. Jain, Phys. Rev. B {\bf 41}, 7653 (1990) and references 
therein.  
\bibitem{KW}   G. Kirczenow, Phys. Rev. B{\bf 53}, 15767(1996).
\bibitem{AB}   J. D. F. Franklin et al., Surf. Sci. 361, 17(1996).
\bibitem{GM}   V. J. Goldman et al., Phys. Rev. B{\bf 55}, 4081(1997). 
\bibitem{YZ1}  Y. Yu, Z. Y. Zhu, Preprint cond-mat/9704124.
\bibitem{ZWJ}  W. J. Zheng, to be published in Int. J. Mod. Phys. B.
\bibitem{PM}   Our work is complementary to the numerical efforts to justify CLL
 microscopically, see for example J. J. Palacios, A. H. Macdonald, Phys. Rev. Lett.{\bf 76}, 118(1996).
\bibitem{CS}F. Calogero, J. Math. Phys. {\bf 10}, 2197 (1967).
\bibitem{IGN}  We therefore ignore the residue interactions in this letter because of their
 irrelevance to the critical properties of the present edge model.            
\bibitem{MEC}  We stress that the crossover behavior between FL and LL is merely a finite 
temperature effect, while the ground state properties at $T=0$ of our edge model(CSM)
are definitely of LL nature.
\bibitem{YY}   C. N. Yang and C. P. Yang, J. Math. Phys. {\bf 10}, 1115(1969).
\bibitem{Wu}   Y. S. Wu and Y. Yu, Phys. Rev. Lett. {\bf 75}, 890 (1995).
\bibitem{KF}   C. L. Kane and M. P. A. Fisher, Phys. Rev. B{\bf 46}, 15233(1992)
; Phys. Rev. Lett{\bf 68}, 1220{1992}.
\bibitem{CPW}  A. M. Chang, L. N. Pfeiffer and K. W. West, Phys. Rev. Lett{\bf77},
2538(1996).
\bibitem{EXP3} The variation between samples should be irrelevant, because the difference 
between 1.75 and 1.5 is large enough to deserve intrinsic explanation.
\bibitem{CO}   For simplicity in numerical calculations, we ignore the contributions
 of the crossover region.
\bibitem{ZM}   The corrections to $N(E)$ due to other factors such as the long
range Coulomb interaction are possible, whose effects on the tunneling conductance
 remain to be verified, see U. Zulicke, A. H. Macdonald, Bull. Am. Phys. Soc.
{\bf 41}, 1, 79(1996).
\bibitem{ZY}   W. J. Zheng and Y. Yu, in preparation.
\bibitem{EXP2} Strictly speaking, TBA is exact only for integrable system like 
the CSM. The addition of Coulomb interaction here is treated perturbatively as
an extension of TBA.
\end{references}
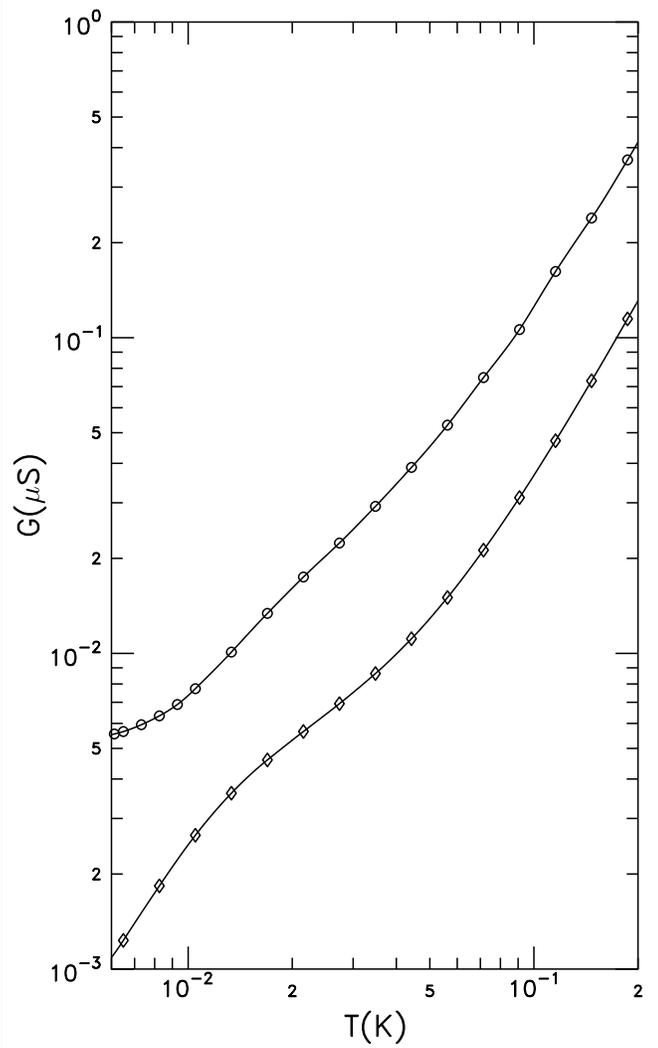
\end{document}